\documentclass[review]{elsarticle}
\usepackage{caption}
\usepackage{subcaption}
\usepackage{lineno,hyperref}
\usepackage{amsmath}
\usepackage{amssymb}
\usepackage{multirow}
\usepackage{svg}
\usepackage{siunitx}
\usepackage{lineno}
\journal{NIM - A}

\begin{document}

\begin{frontmatter}

\title{Simulation study on the optical processes at deep-sea neutrino telescope sites}

\author[pku]{Fan Hu}
\author[sjtu]{Zhenyu Wei}
\author[tdli]{Wei Tian}
\author[tdli]{Ziping Ye}
\author[tdli]{Fuyudi Zhang}
\author[sjtu]{Zhengyang Sun}
\author[sjtu]{Wei Zhi}
\author[sjtu]{Qichao Chang}
\author[sjtu]{Qiao Xue}
\author[pku,kiaa]{Zhuo Li}
\author[tdli,sjtu]{Donglian Xu \corref{corresponding}}
\ead{donglianxu@sjtu.edu.cn}

\cortext[corresponding]{Corresponding author}
\address[pku]{Department of Astronomy, School of Physics, Peking University, 5 Yiheyuan Rd, Beijing, China}
\address[tdli]{Tsung-Dao Lee Institute, Shanghai Jiao Tong University, 520 Shengrong Rd, Shanghai, China}
\address[kiaa]{The Kavli Institute for Astronomy and Astrophysics, Peking University, 5 Yiheyuan Rd, Beijing, China}
\address[sjtu]{School of Physics and Astronomy, Shanghai Jiao Tong University, 800 Dongchuan Rd, Shanghai, China}

\begin{abstract}
The performance of a large-scale water Cherenkov neutrino telescope relies heavily on the transparency of the surrounding water, quantified by its level of light absorption and scattering.
A pathfinder experiment was carried out to measure the optical properties of deep seawater in South China Sea with light-emitting diodes (LEDs) as light sources, photon multiplier tubes (PMTs) and cameras as photon sensors. 
Here, we present an optical simulation program employing the \textsf{Geant4} toolkit to understand the absorption and scattering processes in the deep seawater, which helps to extract the underlying optical properties from the experimental data. 
The simulation results are compared with the experimental data and show good agreements.
We also verify the analysis methods that utilize various observables of the PMTs and the cameras with this simulation program, which can be easily adapted by other neutrino telescope pathfinder experiments and future large-scale detectors.

\end{abstract}

% \begin{highlights}
% \item Research highlight 1
% \item Research highlight 2
% \end{highlights}


\end{frontmatter}

% \linenumbers

\section{Introduction}
\label{sec:intro}

Neutrino telescopes observe neutrinos by detecting the faint Cherenkov light emitted by the secondary charged particles from neutrino-nucleon interactions \cite{AMANDA_overview:2000, ANTARES_overview:2011, icecube_instrument_review:2010, KM3Net_letter_of_intent:2016, Baikal_status:2011}. 
The optical properties of a medium, known primarily as absorption and scattering, determine the propagation behavior of the Cherenkov light, which are critical for optimally improving the geometry layout and angular resolution of a neutrino telescope \cite{Wiebusch:2003, IceCube_source_sensitivity:2003, IceCube-Gen2_layout:2021}.
Therefore, measuring the optical properties of the medium at telescope sites \textit{in-situ} is indispensable for designing and operating neutrino telescopes \cite{ANTARES_measurement:2004, IceCube_measurement:2006, NEMO_measurement:2006, Grace_measurement:2011, Baikal_measurement:2012, Grace_measurement:2018, P-One_measurement:2021}.

The tRopIcal DEep-sea Neutrino Telescope (TRIDENT) pathfinder experiment  (TRIDENT EXplorer, T-REX for short) \cite{TRIDENT_arxiv:2022} measures the optical properties at the South China Sea, which will be the location of the future TRIDENT neutrino telescope. 
The experimental facility is composed of one light emitter module placed in the middle and two light receiver modules separated aside at distances of $21.73\pm0.02\,\mathrm{m}$ and $41.79\pm0.04\,\mathrm{m}$ from the emitter, respectively. 
The emitter module contains LEDs that emit light in both steady and pulsing modes. Photons emitted from the LEDs are diffused by a double-layer diffuser and are isotropic in the emitting region \cite{Wenlian:2022}. 
They will propagate through the vast deep seawater, and a small fraction of them will register at the receiver modules, observed by the cameras or the PMTs.
When the emitter module works in the steady mode, cameras in the receiver modules take photographs of the emitter sphere to obtain the spatial distribution of the received photons. 
While in the pulsing mode, the PMTs inside the receiver modules record the arrival time of the photons, which may be delayed due the scattering processes. 
A photon arrival time distribution (ATD) is built from repeating millions of \textit{in-situ} pulsing experiments.
The images from the cameras and the ATD from the PMTs are then used to extract the optical properties independently, providing complementary verification of the final experimental results. 
Furthermore, the two emitter-receiver pairs with different distances setup  enables a relative measurement to the optical properties of the deep seawater, which largely cancels out the hidden systematic errors.

Since a photon's kinematic state (position, direction, and propagation time) is hard to predict analytically after scattering for a few times, a numerical simulation is required for an in-depth study to accurately interpret the experimental results. 
In this work, we have developed a simulation program to replicate the T-REX apparatus setup and produce photon propagation experiments with various optical property parameters in the deep seawater. 
The simulation output are then used to reconstruct the experimental observables and verify the analysis methods. 

The paper is organized as follows: in Sec. \ref{sec:optical_physics} we briefly overview the optical physics studied. In Sec. \ref{sec:program}, a simulation program based on \textsf{Geant4} \cite{GEANT4:2003, GEANT4:2006} that simulates the photon propagation in the medium is presented. Then, the reconstruction of experimental observables using the simulation output is shown in Sec. \ref{sec:simulation_results}. The validation of the methods to extract optical properties from experimental observables with the help of simulation are studied in detail in Sec. \ref{sec:validation_of_analysis_methods}. A discussion about the effects of geometry simplification and the characteristics of the two systems is performed in Sec. \ref{sec:discussion}, and Sec. \ref{sec:summary} summarizes the main results of the paper.

\section{Optical physics}
\label{sec:optical_physics}

Electromagnetic waves propagating in a medium deposit energy and excite the electrons, atoms, molecules, or solid particles in the medium \cite{book_for_optical:1998}. Upon excitement, the deposited energy can either be converted to thermal energy, in a process called \textit{absorption}, or produce secondary radiation, known as \textit{elastic scattering} (named in this work simply as \textit{scattering}, unless explicitly stated otherwise) if the radiation remains the same frequency. Though the above phenomenon is studied and understood in terms of electromagnetic waves, it can be used to describe the propagation behavior of photons in the quantum world according to the wave-particle dualism nature of electromagnetic waves. 

The scattering processes can be qualitatively classified into two regions with respect to the relative size between the obstacles and the photon wavelength: Rayleigh scattering and Mie scattering. 
Rayleigh scattering refers to the scattering from small obstacles with radius $r \ll \lambda$, where $\lambda$ is the photon wavelength. 
On the other hand, Mie scattering describes the scattering from obstacles with radius $r \gtrsim \lambda$. 
The Mie scattering is mainly caused by the organics in water and usually are the dominant process in natural water \cite{ANTARES_measurement:2004, Baikal_measurement:2012}. 

The angular distribution of scattered photons can be described by a phase function $\beta(\cos\theta)$, with $\theta$ being the photon deflection angle.
For Rayleigh scattering, $\beta_\mathrm{Ray}(\cos\theta)$ can be described by a dipole radiation model and is symmetric over the polarization direction of the incident photons. 
The Mie scattering is much more complex due to the non-negligible size of the obstacles. 
Its phase function $\beta_\mathrm{Mie}(\cos\theta)$ has a strong peak in the forward direction and can be parameterized by the Henyey-Greenstein (HG for short) approximation \cite{Henyey_Greenstein:1941}. 
In summary, the pahse functions for Rayleigh and Mie scatterings can be given, respectively, by:
\begin{linenomath*}
\begin{subequations}
    \label{eq:phase_function}
    \begin{align}
        \beta_\mathrm{Ray}(\cos\theta) &= \frac{3}{8} \left( 1 + \cos{\theta}^2 \right) \label{eq:phase_ray}  \\
        \beta_\mathrm{Mie}(\cos\theta) &= \frac{1}{2} \frac{1 - \mu^2}{(1 + \mu^2 - 2 \mu \cos{\theta}) ^{\frac{3}{2}}} , \label{eq:phase_mie}
    \end{align}
\end{subequations}
\end{linenomath*}
where $\mu$ is the only parameter in the HG formula and has the same value as the expectation of $\cos\theta$.

The probabilities of photons encounter absorption, Rayleigh scattering, and Mie scattering are characterized by the mean free paths and are marked as $\lambda_\mathrm{abs}$, $\lambda_\mathrm{Ray}$, and $\lambda_\mathrm{Mie}$ respectively. There exist many quantities constructed to describe the combined effects of absorption and scattering. The most widely known one is the attenuation length, defined as: 
\begin{linenomath*}
\begin{equation}
    \frac{1}{\lambda_\mathrm{att}} = \frac{1}{\lambda_\mathrm{abs}} + \frac{1}{\lambda_\mathrm{Ray}} + \frac{1}{\lambda_\mathrm{Mie}} .
    \label{eq:attenuation}
\end{equation}
\end{linenomath*}

The physical meaning is the exponential decay of the radiance (having a unit of energy per time per area per solid angle) of a collimated light beam $I_\mathrm{beam}(d)$ over the distance $d$ when propagating in the medium, since both absorption and scattering can eliminate photons in the light beam. 
The decay of $I_\mathrm{beam}(d)$ is characterized by the Beer-Lambert law \cite{Mayerhofer:2020}:
\begin{linenomath*}
\begin{equation}
    I_\mathrm{beam}(d) = I_\mathrm{beam}(0) e^{-d / \lambda_\mathrm{att}} .
    \label{eq:attenuation_meanning}
\end{equation}
\end{linenomath*}

Another typical quantity used in previous experiments \cite{ANTARES_measurement:2004} is the scattering angle averaged attenuation length, defined as:
\begin{linenomath*}
\begin{equation}
    \frac{1}{\lambda_\mathrm{att,avg}} \equiv \frac{1}{\lambda_\mathrm{abs}} + \frac{1}{\lambda_\mathrm{Ray}} + \frac{1 - \mu}{\lambda_\mathrm{Mie}} ,
    \label{eq:att_avg}
\end{equation}
\end{linenomath*}
The $\lambda_\mathrm{att,avg}$ considers that the strong forward scattering processes have a tiny influence on the overall behavior of a bulk of photons. 
Thus it can describe the overall fade-off of photons in scenarios where the small changes in the photon direction can be ignored.
The effects of the physical quantities mentioned above will be further discussed by dedicated simulation studies in Sec. \ref{sec:simulation_results} and \ref{sec:validation_of_analysis_methods}

It should be noted that there exist other optical processes, such as inelastic scattering \cite{Vountas:2003} in deep seawater. 
Since their probabilities of occurrence are orders of magnitude smaller than absorption and elastic scattering, they can be safely ignored in our study.

\section{Simulation program}
\label{sec:program}

\subsection{Set up simulation program using Geant4}

\begin{figure}[!ht]
    \centering
    \includegraphics[width=1.0\linewidth]{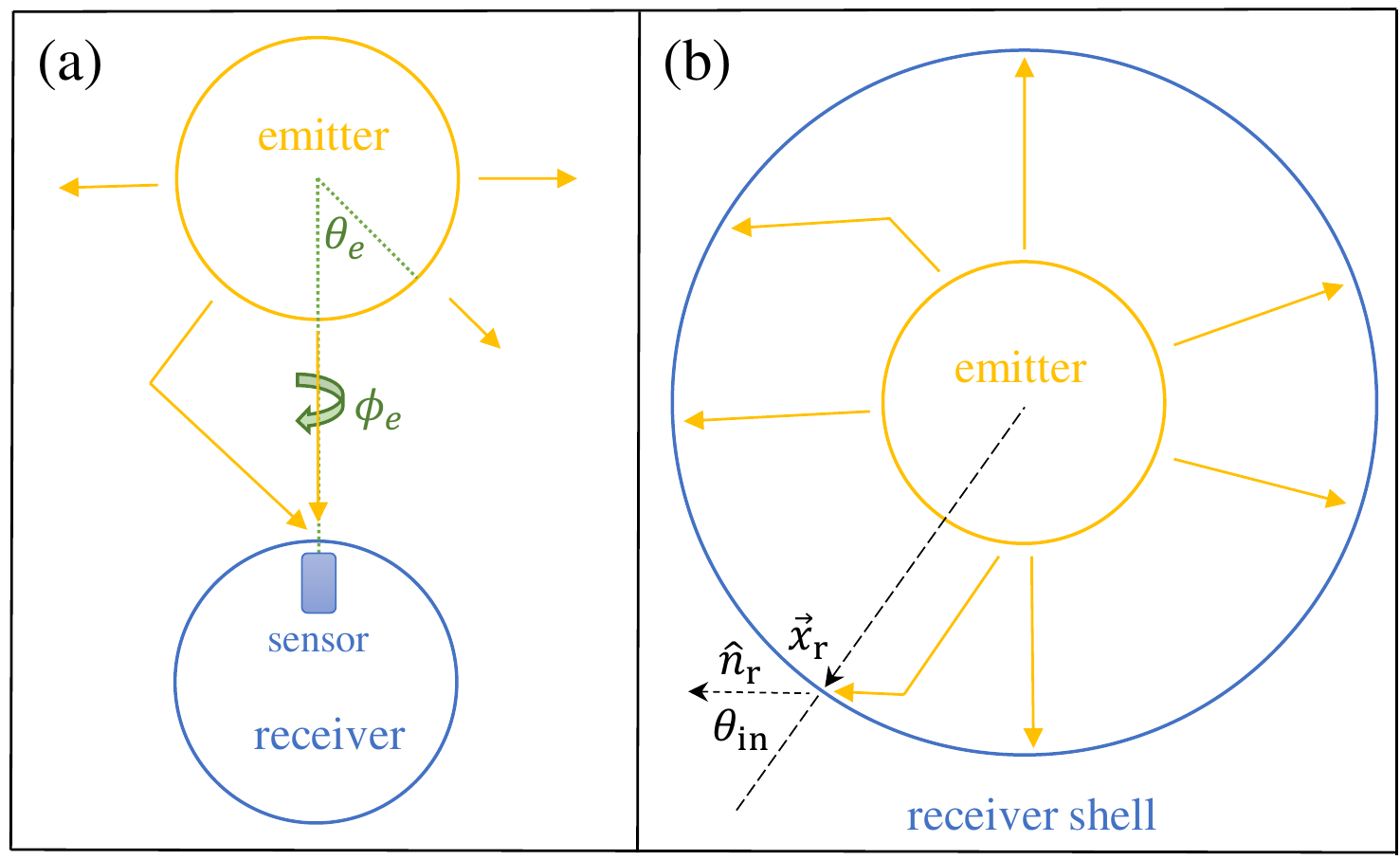}
    \caption{Illustration of the geometry for T-REX (left) and in simulation (right). The receiver module is extended to a receiver shell surrounding the emitter module to increase the simulation efficiency. The yellow lines indicate possible photon trajectories. $\vec{x}_r$ and $\hat{n}_r$ are photon hit position and direction on the shell, $\theta_\mathrm{in}$ is the incident angle. }
    \label{fig:geometry}
\end{figure}

The simulation program \footnote{https://github.com/TRIDENT-Neutrino-Telescope/Pathfinder-Optical-Simulation} is built with \textsf{Geant4} toolkit, a framework widely used in particle physics community \cite{KM3Net_mDOM_simulation:2016, IceCube-Gen2_sensor:2021, JUNO_simulation:2018}. 
\textsf{Geant4} performs particle simulation in a particle-by-particle and step-by-step manner. 
Its modularized design helps users to easily apply the physics processes and extract information from each step.

In this simulation, we assume photons are sufficiently diffused inside the light emitter module. 
Thus photons are emitted isotropically from the surface of the module, which means the radiant intensity follows Lambert's cosine law. 
Once emitted, they propagate through the optical medium and may encounter scattering and absorption processes. 
The optical physics described above has already been implemented in the context of \textsf{G4OpticalPhysics} modules in \textsf{Geant4} and can be applied directly in simulation. 
\textsf{Geant4} can trace every photon till it is absorbed or arrives at the detector. If a photon successfully arrives at the detector, its  kinematic status will be recorded by the output module.

The main drawback of \textsf{Geant4} might be its performance in speed when tracing billions of photons. 
This can be eased by extending the photon detection area in simulation. 
For T-REX, the photons are emitted and diffused by the double-layer diffusers inside the emitter module. 
After emitting from the module, those photons propagate nearly isotropically in the vast seawater environment. 
Only a small portion of photons can eventually hit the photon sensors, as shown in the left panel of Fig. \ref{fig:geometry}. 
Given that the optical properties of deep seawater are isotropic, and assuming the photons from the light emitter are isotropic as well, a photon receiver located at different directions from the emitter should observe consistent results.
In the simulation, we utilize such symmetry and extend the receiver module to a receiver shell surrounding the emitter module.
The simplified geometry is illustrated in the right panel of Fig. \ref{fig:geometry}.
Such adjustment can efficiently boost the number of received photons in the simulation and shorten the simulation time. 
It should be emphasized that while the light emitting module is not perfectly isotropic in spherical space, the in-homogeneous effect is calculated by adding a weighting term in the final result, which will be further described in Sec. \ref{sec:hit_time_distribution}.

In constructing the simulation, we fixed the radius of the central emitter module to be $20\,\mathrm{cm}$, which corresponds to the radius of the outer diffuser layer. 
The refractive index of the deep seawater is set to be 1.34 and works as a time scaling factor in the simulation.
Its effect is perturbatively tunable in the data analysis via rescaling of the photon arrival times.
The other optical parameters as well as the distance between the light source and receiver are the input parameters of the simulation and are set in run time.

\subsection{Simulation input and output}

The flexibility of \textsf{Geant4} framework allows us to explore various effects of optical properties on the photon propagation process in the medium.

In the simulation input, we set the distance between the light source and receiver and the optical property parameters of the medium. 
The program can take input from a human-readable configure file in \textsf{YAML} format or command line arguments. 
The kinematic information of a photon hit is recorded in a per-photon manner as simulation output, which is recorded in a ROOT file \cite{ROOT:2011}. 
Table \ref{tab:simulation_io} summarizes the input and output of the simulation.

\begin{table}[]
    \centering
    \begin{tabular}{|c|c|c|}
        \hline
         & Variable name & Physical description \\
        \hline
        \multirow{5}{3em}{Input} & $\lambda_\mathrm{abs}$ & absorption length  \\
        & $\lambda_\mathrm{Ray}$ & Rayleigh scattering length \\
        & $\lambda_\mathrm{Mie}$ & Mie scattering length \\
        & $\mu$ & the parameter in the HG Mie scattering formula \\
        & $d$ & distance between source and receiver \\
        \hline
        \multirow{5}{3em}{Output} & $\vec{x}_\mathrm{e}$ & emission position at source shell  \\
        & $\vec{x}_\mathrm{r}$ & hit position at observation shell \\
        & $\hat{n}_\mathrm{r}$ & photon direction at observation shell \\
        & $t$ & hit time at the observation shell \\
        & $k$ & number of times that photon get scattered \\
        \hline
    \end{tabular}
    \caption{Summary of the simulation input and output.}
    \label{tab:simulation_io}
\end{table}

\section{Construction of experimental observables} \label{sec:simulation_results}

In this section, we show how the experimental observables can be reconstructed by using the kinematic information of photons from the simulation output. 
We launched two simulation runs with emitter-receiver distances of 20\,m and 40\,m, respectively. 
The optical parameter settings used here are summarized in Table \ref{tab:sim_set}, which are representative of the deep seawater. \cite{ANTARES_measurement:2004, P-One_measurement:2021}. 
In each run, $10^7$ photons are emitted from the light emitter module. 

\begin{table}
    \centering
    \begin{tabular}{|c|c|c|c|c|}
        \hline
        \textbf{Parameters} & $\lambda_\mathrm{a}$ & $\lambda_\mathrm{Ray}$ & $\lambda_\mathrm{Mie}$ & $\mu$\\
        \hline
        \textbf{Values} & 27\,m & 200\,m & 70\,m & 0.97\\
        \hline
    \end{tabular}
    \caption{Summary of the default optical parameter settings in simulation. Those parameters are for clean seawater. \cite{ANTARES_measurement:2004, P-One_measurement:2021, TRIDENT_arxiv:2022}}
    \label{tab:sim_set}
\end{table}

\subsection{Camera images}

\begin{figure}[htp]
    \centering
    \includegraphics[width=0.6\textwidth]{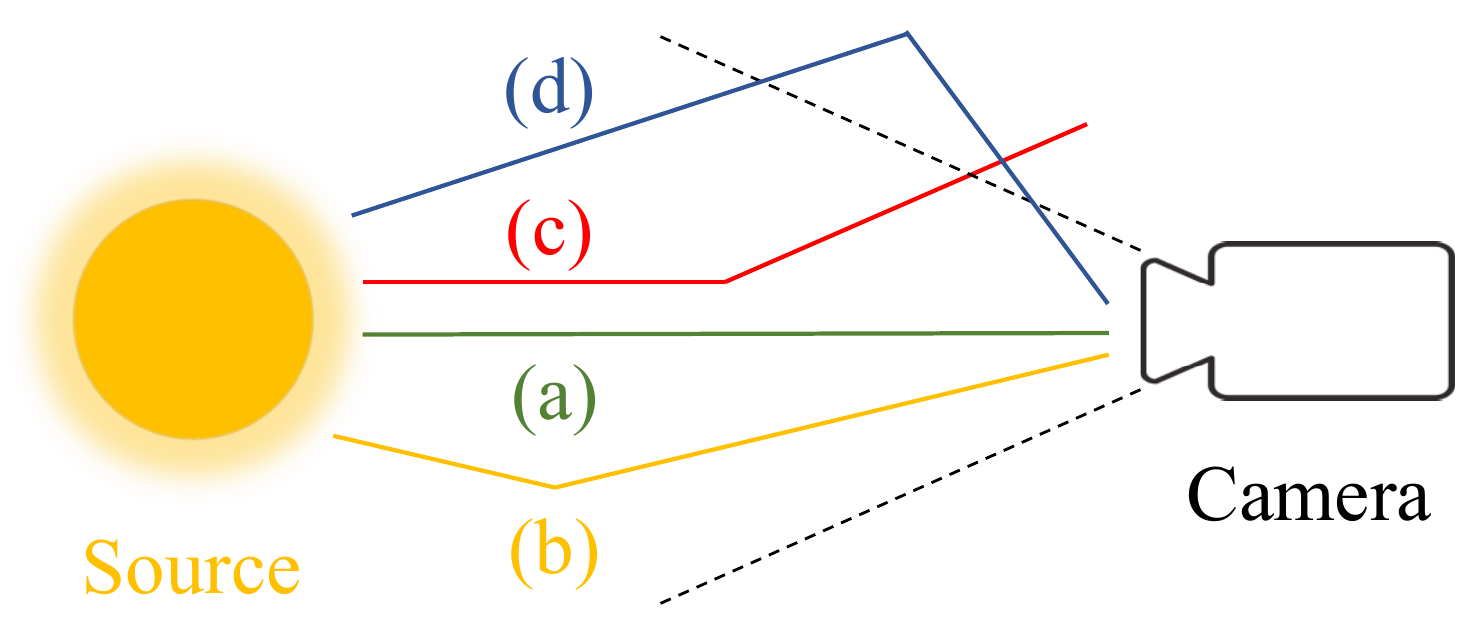}
    
    \centering
    \includegraphics[width=0.5\textwidth]{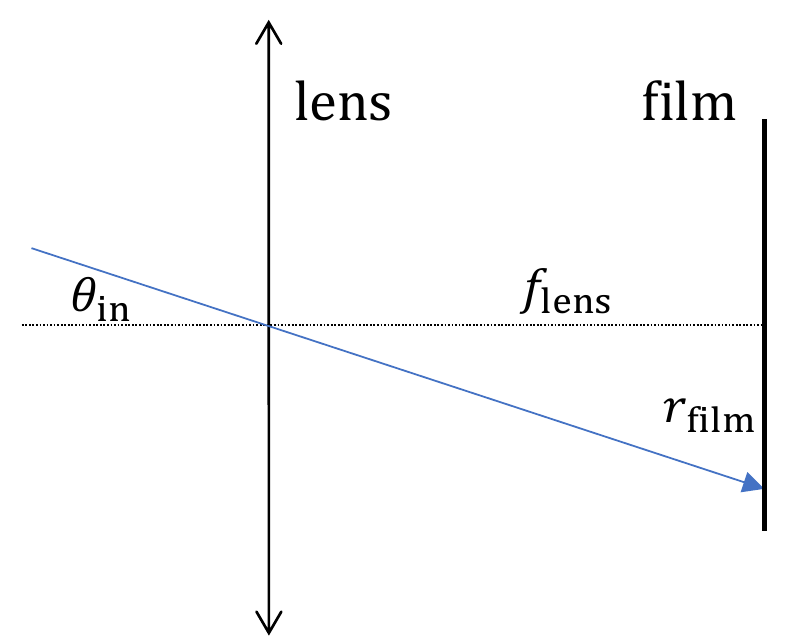}
    \caption{Upper: illustration for the possible trajectories of photons emitted from the light source. The dashed line represents the viewing angle of the camera's field of view. Trajectories illustrated are: (a) directly arriving photon, (b) photon that is scattered into the camera, (c) photon that gets scattered out, (d) photon that is scattered into the camera but is outside the field of view. Among them, only (a) and (b) can be recorded by the camera. Lower: geometry of pinhole camera model. All photons are focused on the CMOS film, and no bokeh effect is simulated. The hit radius at film $r_\mathrm{film}$ corresponds to the incident angle of photon $\theta_\mathrm{in}$, as previously shown in Fig. \ref{fig:geometry}. }
    \label{fig:camera_model}
\end{figure}

In T-REX, a camera is housed in the light receiver module and takes photographs for the light emitter module when it is glowing in steady mode \cite{TianWei:2022}. Once emitted from the source, photons can go through different trajectories, as shown in the upper panel of Fig. \ref{fig:camera_model}, and some of them arrive at the camera and may be recorded by the CMOS film of the camera to form an image. 

In our simulation, we can also construct the image by using the photon direction $\hat{n}_\mathrm{r}$ and position $\vec{x}_\mathrm{r}$ at the receiver sphere. This is done according to the pinhole camera model, as abstracted in the lower panel of Fig. \ref{fig:camera_model}. The incident angle of the photon at the lens $\theta_\mathrm{in}$ can be evaluated by the geometry relation:
\begin{linenomath*}
\begin{equation}
    \theta_\mathrm{in} = \arccos \left( \hat{n}_\mathrm{r} \cdot \frac{\vec{x}_\mathrm{r}}{\vert\vec{x}_\mathrm{r}\vert} \right) . \label{eq:camera_angle}
\end{equation}
\end{linenomath*}
And the radial coordinate of the photon hit point at the camera film $r_\mathrm{film}$ can be derived as:
\begin{linenomath*}
\begin{equation}
    r_\mathrm{film} = f_\mathrm{lens} \sin{\theta_\mathrm{in}} , \label{eq:camera_radius}
\end{equation}
\end{linenomath*}
where we have used $f_\mathrm{lens}$ as the focus of the lens. For the case in T-REX, $f_\mathrm{lens} = 25~\mathrm{mm}$ and the pixel size in CMOS is $3.45~\mathrm{\mu m}$. Thus one pixel corresponds to an incident angle of $1.38 \times 10^{-4} ~\mathrm{rad}$ \cite{TianWei:2022}. 

\begin{figure}[htb]
    \centering
    \includegraphics[width=0.75\linewidth]{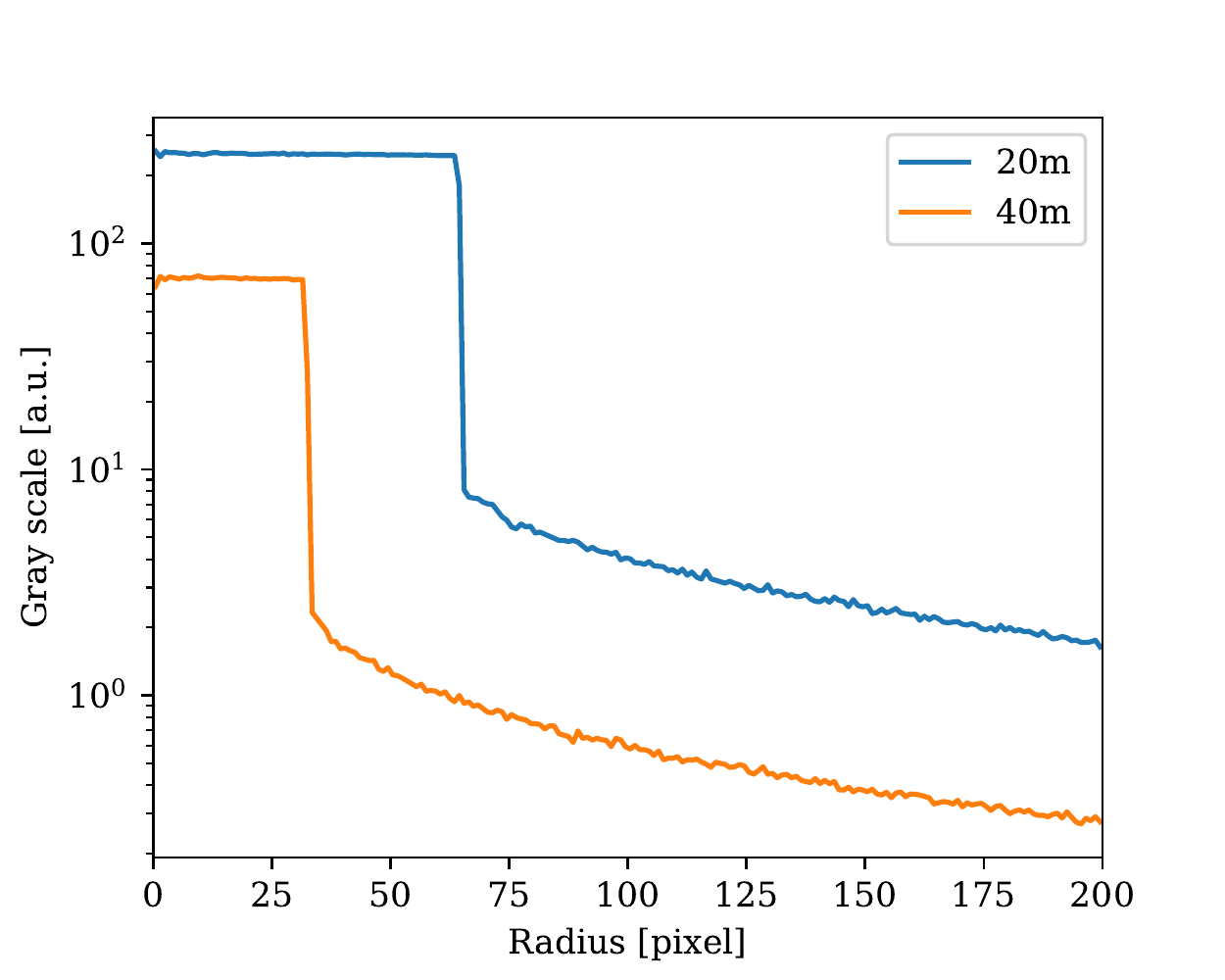}
    \caption{Gray value distribution. The central plateau in the gray value curve corresponds to the image of the light emitter module, while the tail corresponds to the halo caused by the scattering processes.}
    \label{fig:gray_distribution}
\end{figure}

The distribution of radius of hit points at film $r_\mathrm{film}$ is shown in Fig. \ref{fig:gray_distribution}, with two emitter-receiver distances. The linearity of the CMOS response allows us to convert the number of hits into the gray value of images. The central plateau in the gray value curve corresponds to the image of the light emitter module. 

\begin{figure}[htb]
    \centering
    \includegraphics[width=0.95\linewidth]{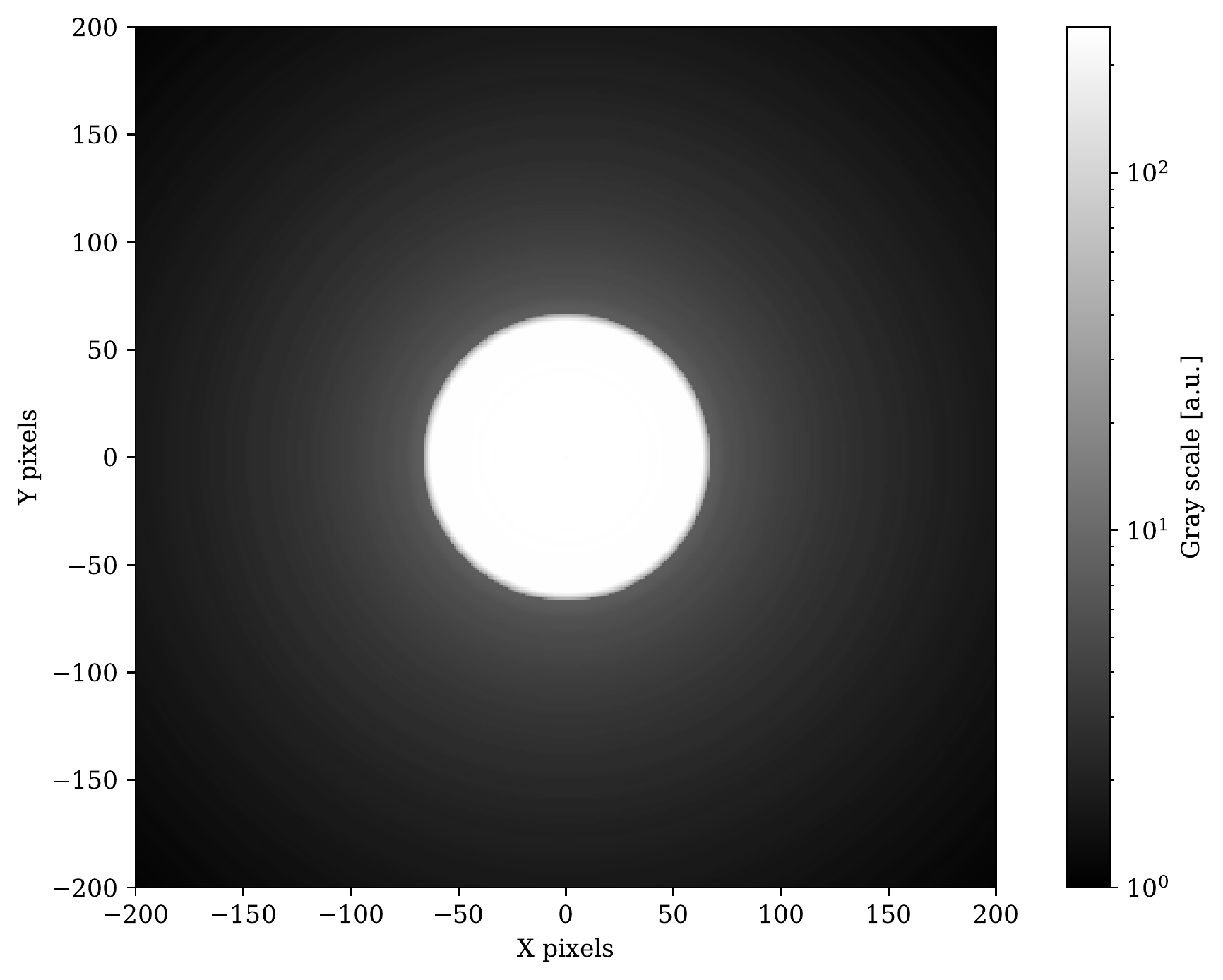}
    \caption{Camera image from simulation. A pinhole camera is taking photons for the light source at a distance of 20\,m. The gray value is shown in log scale to illustrate the gradient of the halo.}
    \label{fig:camera_image}
\end{figure}

Assuming the light source is symmetric in the axial direction, we generate the camera image according to the distribution of $r_\mathrm{film}$. 
Figure \ref{fig:camera_image} shows an example of the resulting image, which consists of a bright central sphere with a surrounding halo of scattered light. 
The halo is caused by scattering light and has a gray value of $\sim 2\%$ compared to the bright central sphere. 

\subsection{Photon arrival time distribution in PMT}
\label{sec:hit_time_distribution}

\begin{figure}[htp]
    \centering
    \includegraphics[width=1.0\linewidth]{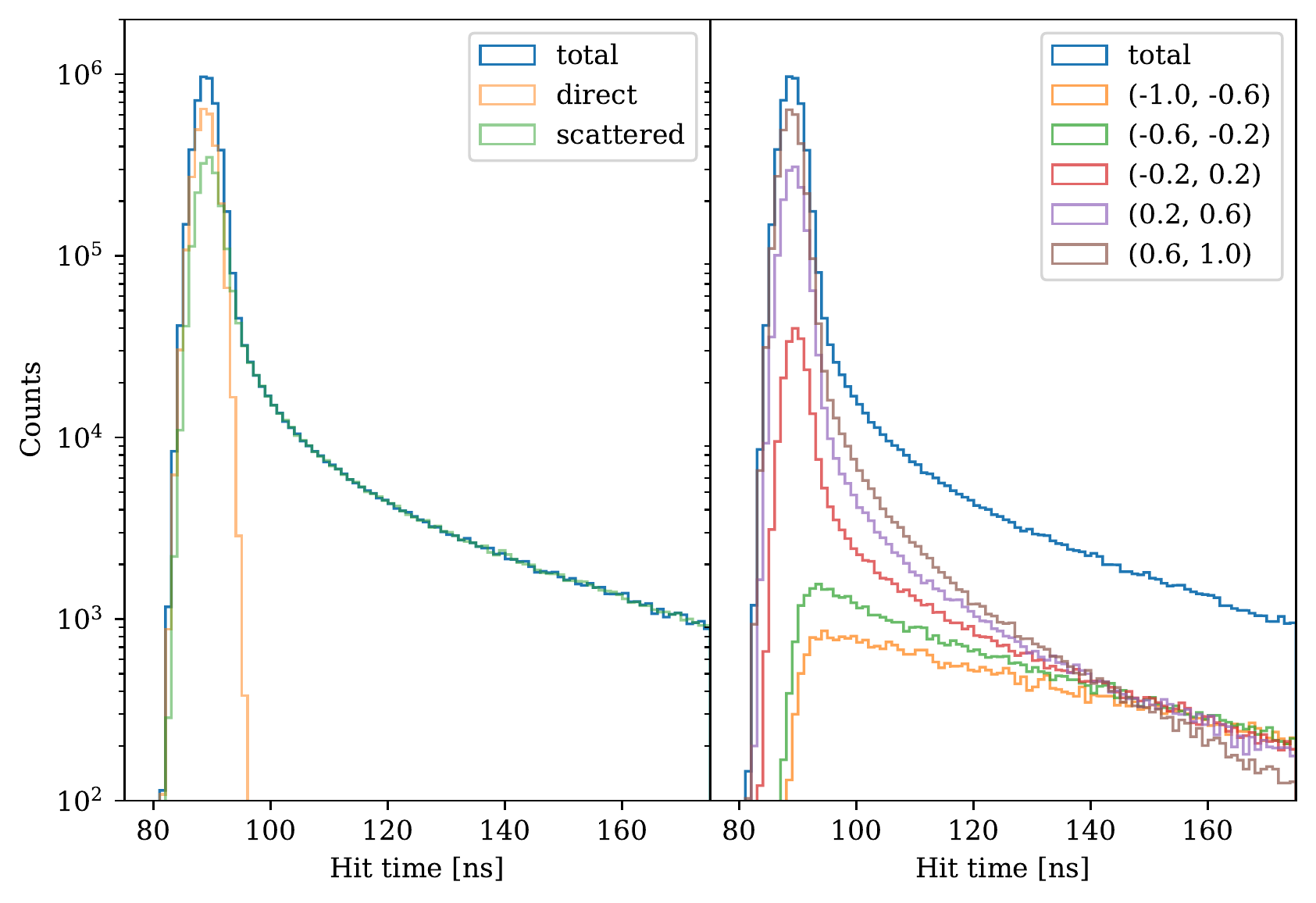}
    \caption{Photon arrival time distribution (ATD) measured in PMT at a distance of 20\,m. A time smearing with FWHM of $4 ~\mathrm{ns}$ is added. In the left panel, the photons are divided into scattered and direct groups. A long scattering tail can be observed in the scattered group. In the right panel, the photons are divided into groups according to $\theta_\mathrm{e}$. The values in the legend represents the binning of $\theta_\mathrm{e}$.}
    \label{fig:pmt_time_curve}
\end{figure}

The PMTs in the receiver module measure the ATD of received photons. The photons that have undergone scattering processes arrive late and form a long tail in the distribution. The scattering effect of the medium can be resolved by fitting the shape and amplitude of the tail \cite{ANTARES_measurement:2004, IceCube_measurement:2006}. The ATD can be calculated using the photon hit time $t$ recorded in the simulation output, as shown in Fig. \ref{fig:pmt_time_curve}. Since the PMTs in T-REX have a transient time spread (TTS) of about $4 ~\mathrm{ns}$ \cite{Yudi:2022}, we have convolved the ATD with a Gaussian kernel with full width at half maximum (FWHM) equal to $4 ~\mathrm{ns}$ to simulate the effect of TTS.

The 3D printed supporter inside the emitter module brings asymmetry to the light intensity in latitude direction \cite{Wenlian:2022}. Such an effect can be studied in simulation by deriving the relative latitude of the photon's emitting point compared to the photon's receiving point $\theta_\mathrm{e}$, defined as:
\begin{linenomath*}
\begin{equation}
    \theta_\mathrm{e} = \arccos \left( \frac{\vec{x}_\mathrm{r}}{\vert\vec{x}_\mathrm{r}\vert} \cdot \frac{\vec{x}_\mathrm{e}}{\vert\vec{x}_\mathrm{e}\vert}  \right) .
    \label{eq:latitude}
\end{equation}
\end{linenomath*}
As shown in the right panel of Fig. \ref{fig:pmt_time_curve}, photons from different $\theta_\mathrm{e}$ have distinct ATDs. The photons from the negative $\theta_\mathrm{e}$ area, which face opposite to the PMT, can only arrive at PMT after being scattered. 
Thus the ATD of those photons has no peak at the geometric expected time but a long tail. 
Once we have evaluated $\theta_\mathrm{e}$, we can add weighting terms to the photons from different $\theta_\mathrm{e}$ to simulate the asymmetry effect of the emitter module in the analysis.

\subsection{Comparison with experimental results}

\begin{figure}[ht]
    \centering
    \includegraphics[width=0.8\linewidth]{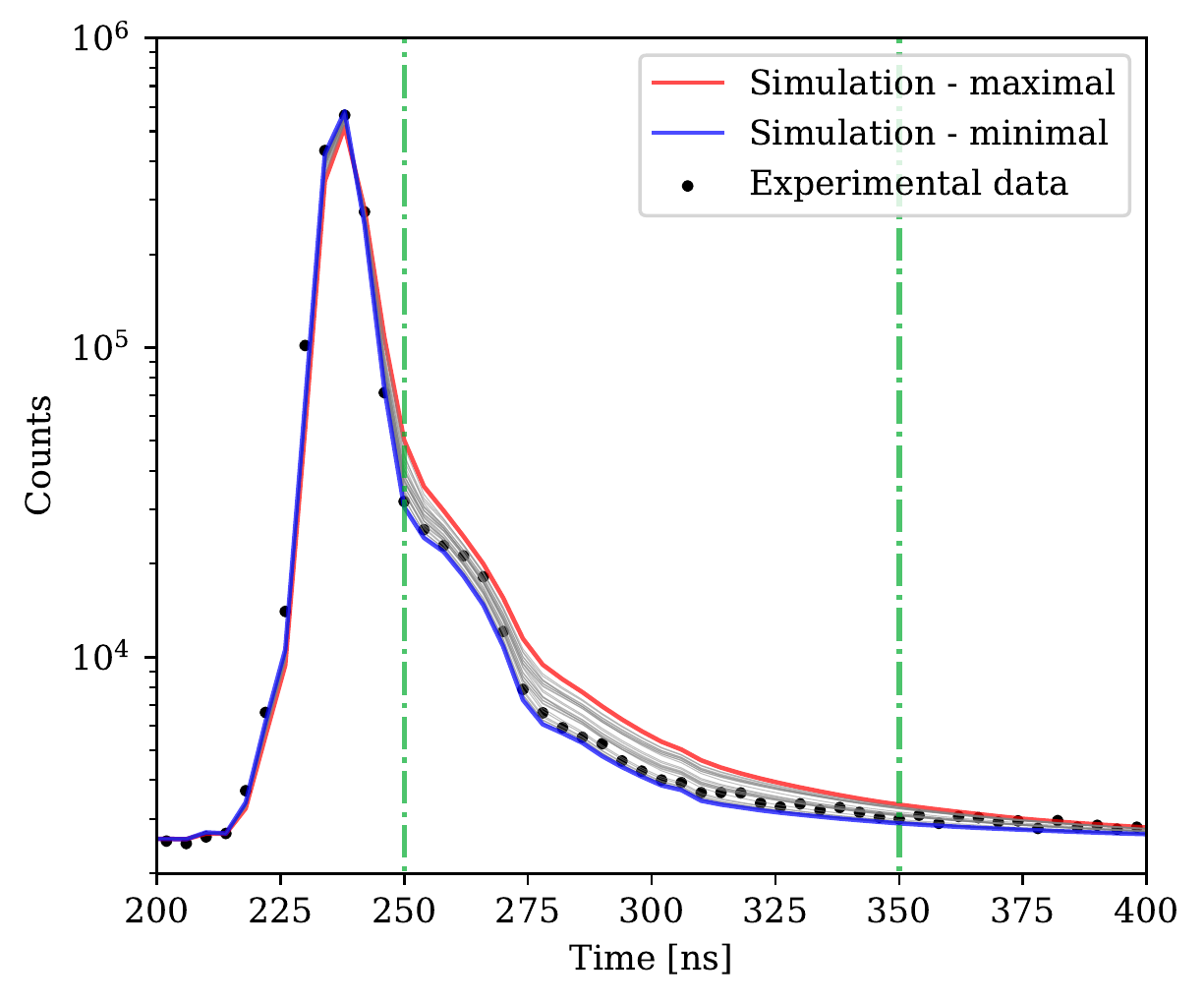}
    \caption{Comparison between the simulation data and experimental data for the photon arrival time distribution (ATD) measured by the PMT system. The black dots correspond to the data taken from the T-REX. The curves correspond to the simulation results after calibration. The simulation input parameters are the combination of: $\lambda_a = 27\,\mathrm{m}$, $\lambda_\mathrm{Ray} = [120\,\mathrm{m}, 180\,\mathrm{m}, 240\,\mathrm{m}]$, $\lambda_\mathrm{Mie} = [30\,\mathrm{m}, 60\,\mathrm{m}, 90\,\mathrm{m}]$ and $\langle \cos\theta \rangle = [0.90, 0.94, 0.98]$, summing up to 27 different configurations. The blue and red curves represent the minimal and maximal scattering cases. The two green dashed lines indicate the dominantly scattering tail region. }
    \label{fig:pmt_exp}
\end{figure}

The simulation results can be compared with the measurements in T-REX after meticulous calibration. 
For the PMT system, the time profile of the pulsing LEDs, the response of PMTs, and readout electronics are calibrated in the laboratory during the T-REX \cite{Yudi:2022}. 
After considering those effects, the simulation result matches well with the experimental data, as shown in Fig. \ref{fig:pmt_exp}. 
The curves from simulation with different optical parameters can cover the dominantly scattering tail in between $250\,\mathrm{ns}$ and $350\,\mathrm{ns}$ observed in the experiment data, where the scattering effect is mostly revealed.

% For the camera system, the profile of images observed by the camera is dominated by the structure of the light emitter module \cite{Wenlian:2022,TianWei:2022}. Thus the comparison between the simulation and experiment is not shown here. But never the less, the simulation is used to guide the analysis method of the camera system, as shown in Sec. \ref{sec:validation_of_analysis_methods}.

A detailed discussion about the $\chi^2$ method that compares the experimental data with calibrated simulation results to obtain the best fit optical properties are presented in \cite{TRIDENT_arxiv:2022}.

% By comparing the gray value distribution shown in Fig. \ref{fig:gray_distribution} and the photon ATD shown in Fig. \ref{fig:pmt_time_curve} with the ones measured in experiment, chi square methods can be built. And the best fit optical properties can be obtained by varying the simulation inputs and searching the minimum of chi square, as performed in \cite{TRIDENT_arxiv:2022}. 

\section{Validation of analysis methods}
\label{sec:validation_of_analysis_methods}

Apart from the brute force $\chi^2$ fitting method mentioned above, the optical properties of the medium can also be evaluated by constructing concise expressions. 
However, theoretical statistic analysis of photons' behavior without low-order approximation appears intricate when both absorption and scattering processes exist and tangle with each other. 
Any experimental macroscopic observable is engendered under the mixture of two distinct optical processes. 
This makes numerical simulation a better way to conduct viable analysis. 
With the help of the simulation program developed above, we can exploit how to measure the optical properties of the medium. 

Note that we are going to construct a variety of \textit{effective} attenuation lengths directly from the simulated observables in the following context. 
Those physical quantities are marked by a hat, e.g., $\hat{\lambda}_\mathrm{att}$, to be distinguished from the canonical quantities defined in Sec. \ref{sec:optical_physics}.

\subsection{Analysis methods for the camera system}

The camera system takes photographs of the steady emitting light source. 
It measures the directions of arrived photons by their positions at the film, as shown in Fig. \ref{fig:camera_model}. 
The gray value of the image observed by the camera system corresponds to the radiance of the light field in physics.

\begin{figure}[htp]
    \centering
    \includegraphics[width=0.65\linewidth]{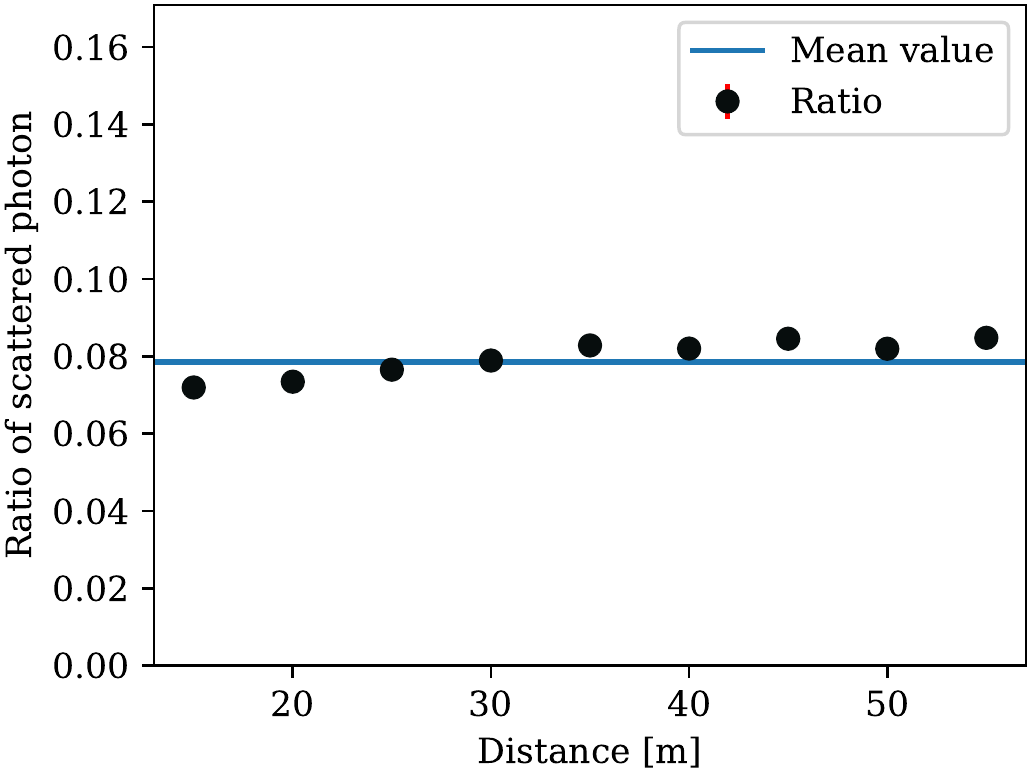}
    \caption{The ratio of scattered photons in the central region of a camera image observed at different distances. It shows that the ratio remains relatively constant over different distances.}
    \label{fig:scattering_ratio}
\end{figure}

We performed numerical simulation using the input optical parameters shown in Table \ref{tab:sim_set} with various distances between the source and receiver, ranging from $15\,\mathrm{m}$ to $55\,\mathrm{m}$, covering the distance range in the experiment. 
It is verified in simulation that the photons composing the central radiance, $I_{\mathrm{center}}$, represented by the gray value of the pixels at the center part of the image (Fig. \ref{fig:gray_distribution}), are mainly directly arriving photons that have not undergone any scatterings during propagation. 
As shown in Fig. \ref{fig:scattering_ratio}, the ratio of the directly arriving photons $\alpha$ in $I_{\mathrm{center}}$ has a stable value of $\alpha = 92 \pm 1 \%$ in the distance range.

\begin{figure}[htp]
    \centering
    \includegraphics[width=0.6\textwidth]{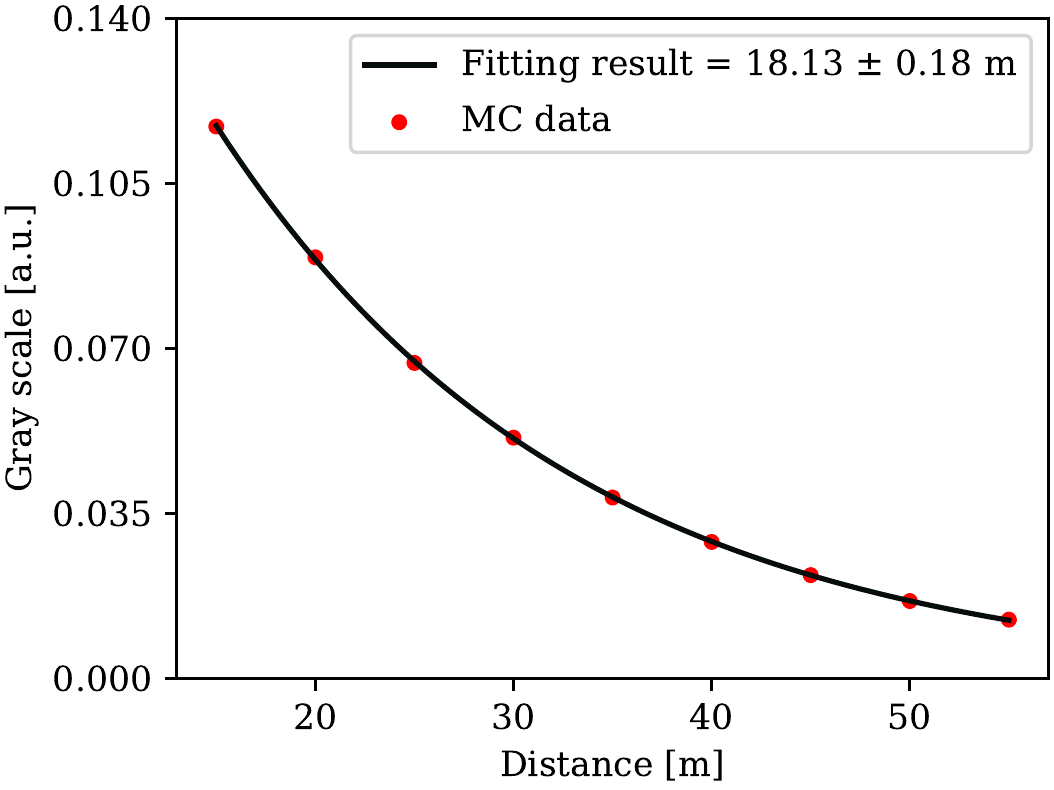}
    \caption{The gray scale of central pixels ($I_\mathrm{center}$) as a function of source-receiver distance. The input optical properties in simulation is summarized in Table \ref{tab:sim_set}, i.e., the attenuation length is $17.75\,\mathrm{m}$. The fitted attenuation length according to Equation \ref{eq:def_ic} is $\hat{\lambda}_\mathrm{att} = 18.13 \pm 0.18 \,\mathrm{m}$, where the error comes from the fitting.}
    \label{fig:camera_algorithm}
\end{figure}

According to the Beer-Lambert law \cite{Mayerhofer:2020}, the attenuation length can be measured by comparing the radiance of directly arriving photons at two distances. 
Here, we use the observed central gray value to fit the attenuation length:
\begin{linenomath*}
\begin{equation}
    I_{\mathrm{center}}(d) \propto \frac{1}{\alpha} \exp(-\frac{d}{\hat{\lambda}_\mathrm{att}}) , 
    \label{eq:def_ic}
\end{equation}
\end{linenomath*}

The verification of this method is shown in Fig. \ref{fig:camera_algorithm}, where we have assumed the ratio of directly arriving photons $\alpha$ is constant over the distance. 
It can be found that fitting result of $\hat{\lambda}_\mathrm{att} = 18.13 \pm 0.18 \,\mathrm{m}$ is close to the canonical attenuation length $\lambda_\mathrm{att} = 17.75 \,\mathrm{m}$. 
The attenuation length obtained using this method is about $2\%$ larger than the canonical attenuation length due to the slightly increasing ratio of scattered photons over the distance. 
The bias of this method is only a minor effect compared to other systematic errors in the measurement of attenuation length in the actual experiment.
If we take this effect into account and select only direct photons in simulation, the fitting result will be $\hat{\lambda}_\mathrm{att} = 17.85 \pm 0.10 \,\mathrm{m}$, which is consistent with the canonical attenuation length.

\subsection{Analysis methods for the PMT system}

\begin{figure}[!ht]
    \centering
    \includegraphics[width=0.8\linewidth]{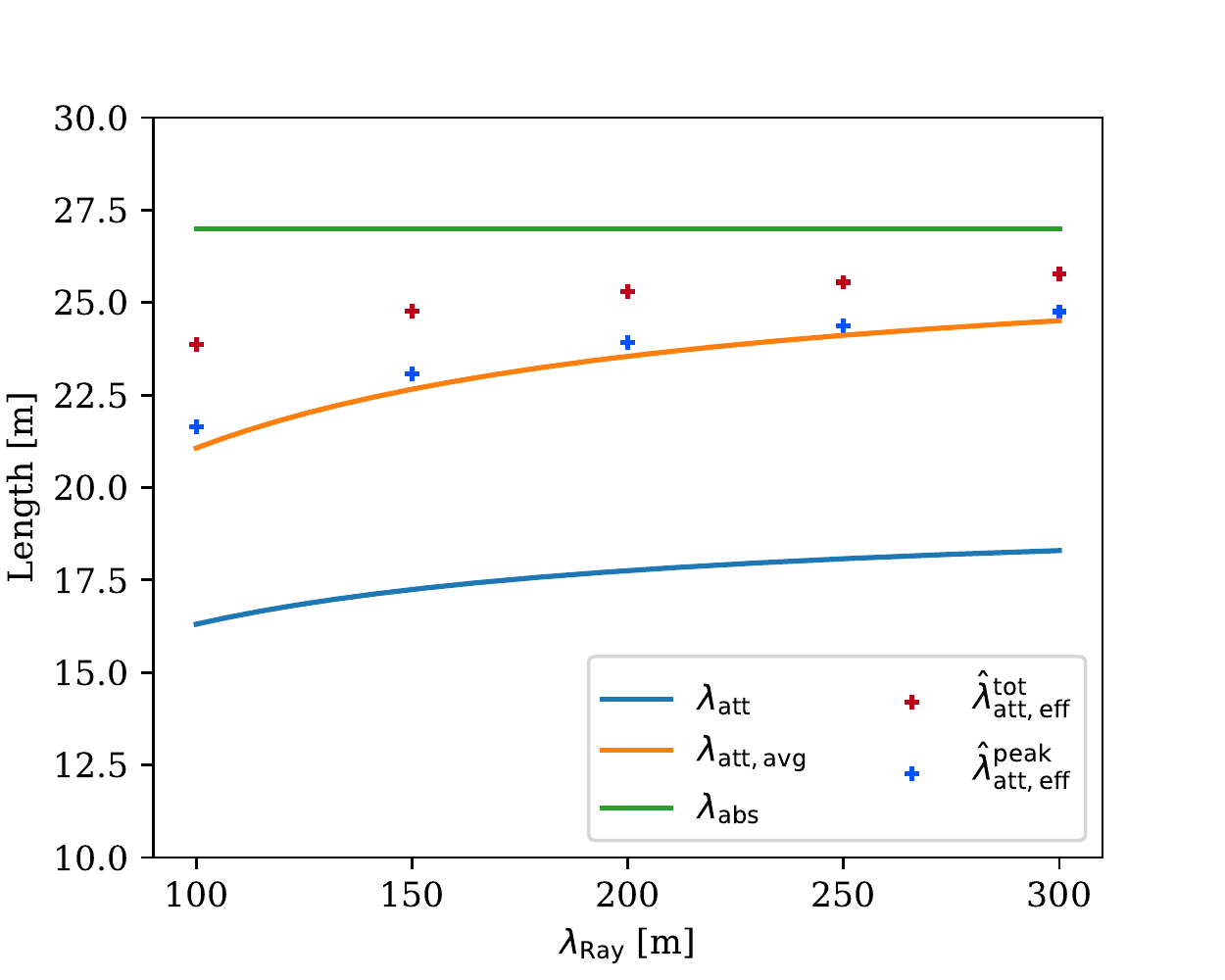}
    \caption{Attenuation lengths with different definitions as a function of Rayleigh scattering length. The optical properties except Rayleigh scattering length are the same as Table \ref{tab:sim_set}. The curves for $\lambda_\mathrm{att}$, $\lambda_\mathrm{att,avg}$ and $\lambda_\mathrm{abs}$ are directly evaluated from simulation truth by their definitions. The red dots represent the measured effective attenuation length using the method described by Equation \ref{eq:att_eff_pmt_tot}. The blue dots are similar to red dots but use the photon numbers from a $3\times$ TTS time window around the peak instead of the whole time window. }
    \label{fig:att_comp}
\end{figure}

\begin{figure}[!ht]
    \centering
    \includegraphics[width=0.55\linewidth]{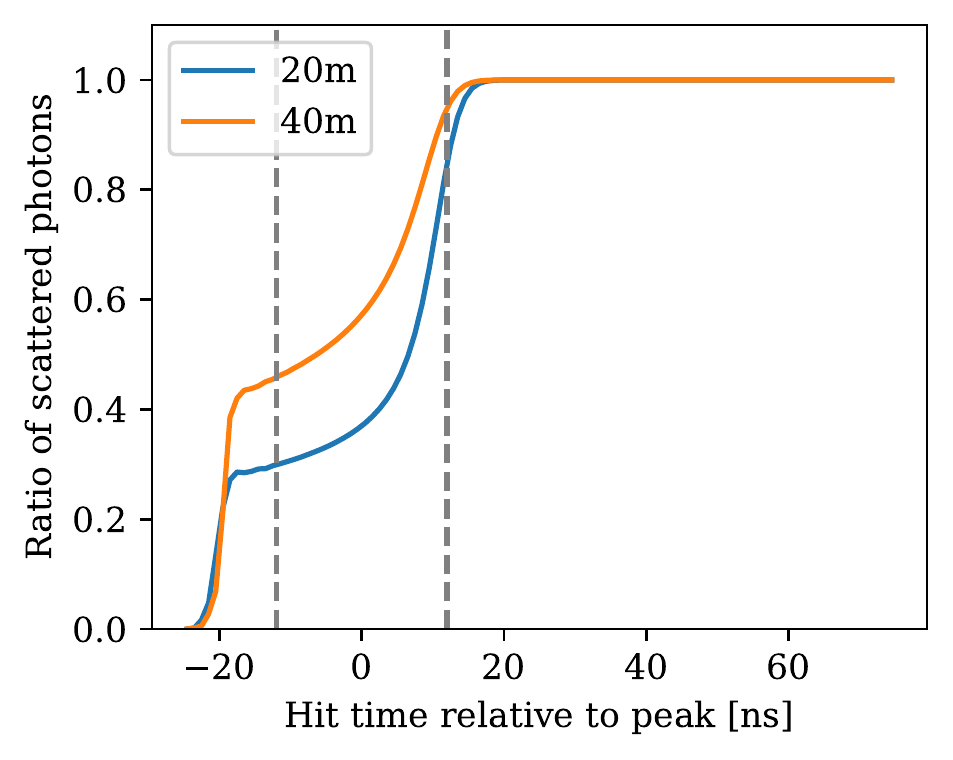}
    \caption{The ratio of scattering photons in each time bin at 20\,m and 40\,m. The dashed line shows the range of integrated bins for the peak-only method: from $-3\times\mathrm{TTS}$ to $+3\times\mathrm{TTS}$. At times after $3\times\mathrm{TTS}$, almost all photons have encountered scattering during propagation. }
    \label{fig:pmt_scatter_ratio}
\end{figure}

The PMT system measures the number of arrived photons as a function of time. 
It can also derive an effective attenuation length $\hat\lambda_\mathrm{att,eff}^\mathrm{tot}$ by comparing the total number of received photons at two distances:
\begin{linenomath*}
\begin{equation}
    \hat{\lambda}_\mathrm{att,eff}^\mathrm{tot} = (d_1 - d_2) / \ln \left( \frac{N_\mathrm{2}^\mathrm{tot} d_2^2}{N_\mathrm{1}^\mathrm{tot} d_1^2} \right) ,
    \label{eq:att_eff_pmt_tot}
\end{equation}
\end{linenomath*}
where $N_\mathrm{1}^\mathrm{tot}$ and $N_\mathrm{2}^\mathrm{tot}$ are the total number of photons received by PMTs at distances of $d_1 = 20\,m$ and $d_2 = 40\,m$, respectively. 

Some previous studies \cite{ANTARES_measurement:2004, P-One_measurement:2021} use $\hat\lambda_\mathrm{att,eff}^\mathrm{tot}$ (Equation \ref{eq:att_eff_pmt_tot}) to estimate the canonical attenuation length $\lambda_\mathrm{att}$ (Equation \ref{eq:attenuation}). 
To verify this method, we performed multiple simulation runs with various Rayleigh scattering lengths $\lambda_\mathrm{Ray}$ and recorded outputs at distances of 20\,m and 40\,m. 
The other optical parameters in simulation remain the same as Table \ref{tab:sim_set}. 
Our simulation shows that their deviation can be as significant as $\sim 40\%$, shown in Fig. \ref{fig:att_comp}. 
The deviation between $\hat\lambda_\mathrm{att,eff} ^\mathrm{tot}$ and $\lambda_\mathrm{att}$ is caused by the large ratio of scattered photons observed by PMTs, which increases with the emitter-receiver distance, as shown in the left panel of Fig. \ref{fig:pmt_time_curve} and Fig. \ref{fig:pmt_scatter_ratio}. 
We find that the ratios of scattered photons among all photons observed by the PMTs are 43\% and 67\% in distances of 20\,m and 40\,m, respectively. 
Since our light source is not beamed to the receiver but is isotropically emitting photons in all directions, some photons that initially emitted towards a different direction to sensor might be received by the PMT, which has a large viewing angle, after encountering multiple times scattering. 
Our study shows that $\hat\lambda_\mathrm{att,eff} ^\mathrm{tot}$ measured by the above method can not be used to estimate the canonical attenuation length of the medium.

By restricting the integration window to $3 \times \mathrm{TTS}$ around the peak of ATD instead of the whole windows, we can derive another experimental observable $\hat{\lambda}_\mathrm{att,eff}^\mathrm{peak}$. 
As shown in Fig. \ref{fig:att_comp}, we find that $\hat{\lambda}_\mathrm{att,eff}^\mathrm{peak}$ has a similar value to the scattering angle averaged attenuation length $\lambda_\mathrm{att, avg}$ (Equation \ref{eq:att_avg}), and their relative error is within $3\%$ in the cases of our input optical parameters (Table \ref{tab:sim_set}). 
As discussed in Sec. \ref{sec:optical_physics}, $\lambda_\mathrm{att, avg}$ is larger than $\lambda_\mathrm{att}$ because it considers the strong forwardness of the deflection angle in Mie scattering. 
In the case of PMT measurement, the photons arriving around the peak time are either non-scattered or slighted scattered, thus their number is approximately corresponding to $\lambda_\mathrm{att, avg}$.

Despite the difficulties in measuring the canonical attenuation length using the PMT system, the absorption length of the medium can be measured well using the timing ability of PMT:
\begin{linenomath*}
\begin{equation}
    \int_{-\infty}^{\infty} \mathcal{N}_1(t) e^{\frac{c t}{n \lambda_\mathrm{abs}}} ~\mathrm{d}t \times d_1^2 = 
    \int_{-\infty}^{\infty} \mathcal{N}_2(t) e^{\frac{c t}{n \lambda_\mathrm{abs}}} ~\mathrm{d}t \times d_2^2 ,
    \label{eq:abs_bin}
\end{equation}
\end{linenomath*}
where $t$ is arrival time of photon hits, $\mathcal{N}(t) = \frac{\mathrm{d}N}{\mathrm{d}t}$ is the ATD (number density per unit time) of photons observed by PMT, $n$ is the refractive index of the medium, $c$ is the speed of the light, and $d$ is the emitter-receiver distance. $i$ indicates the two measurements at a distance of $20\,\mathrm{m}$ and $40\,\mathrm{m}$.
This measurement utilizes the timing ability of PMTs, which can be used to convert the photon arrival time to the trajectory length $l$ of the photons using relation $l = c t / n$. 
The probability of the photon not being absorbed through a trajectory length corresponds to $t$ is $e^{\frac{c t}{n \lambda_\mathrm{abs}}}$ and is used as the weighting term in Equation \ref{eq:abs_bin}.

An advantage of using Equation \ref{eq:abs_bin} is its invariance under convolution. 
Due to the transient time spread of PMT and the LED emitting pulse width, the observed ATD $N_\mathrm{obs}(t)$ is smeared. 
Assuming the smearing kernel function $G(\tau)$ is the same for the two PMTs, it can be proved that such a smearing effect can bring only a form factor $\mathcal{G} = \int_{-\infty}^{\infty} G(\tau) e^{\frac{c\tau}{n\lambda_\mathrm{abs}}} ~\mathrm{d}\tau$:
\begin{linenomath*}
\begin{equation}
    \begin{aligned}
    & ~~~\int_{-\infty}^{\infty} \mathcal{N}_\mathrm{obs}(t) e^{\frac{ct}{n\lambda_\mathrm{abs}}} ~\mathrm{d}t \\
    &= \int_{-\infty}^{\infty} G(\tau) \mathcal{N}(t-\tau) e^{\frac{ct}{n\lambda_\mathrm{abs}}} ~\mathrm{d}t \mathrm{d}\tau \\
    &= \int_{-\infty}^{\infty} G(\tau) e^{\frac{c\tau}{n\lambda_\mathrm{abs}}} ~\mathrm{d}\tau \times
    \int_{-\infty}^{\infty} \mathcal{N}(t-\tau) e^{\frac{c(t-\tau)}{n\lambda_\mathrm{abs}}} ~\mathrm{d}t \\
    &= \mathcal{G} \int_{-\infty}^{\infty} \mathcal{N}(t) e^{\frac{ct}{n\lambda_\mathrm{abs}}} ~\mathrm{d}t
    \label{eq:abs_convolve}
    \end{aligned}
\end{equation}
\end{linenomath*}
Thus Equation \ref{eq:abs_bin} remains valid after replacing the original ATD $\mathcal{N}(t)$ by the smeared distribution $\mathcal{N}_\mathrm{obs}(t)$.

\begin{figure}[htp]
    \centering
    \includegraphics[width=1.0\linewidth]{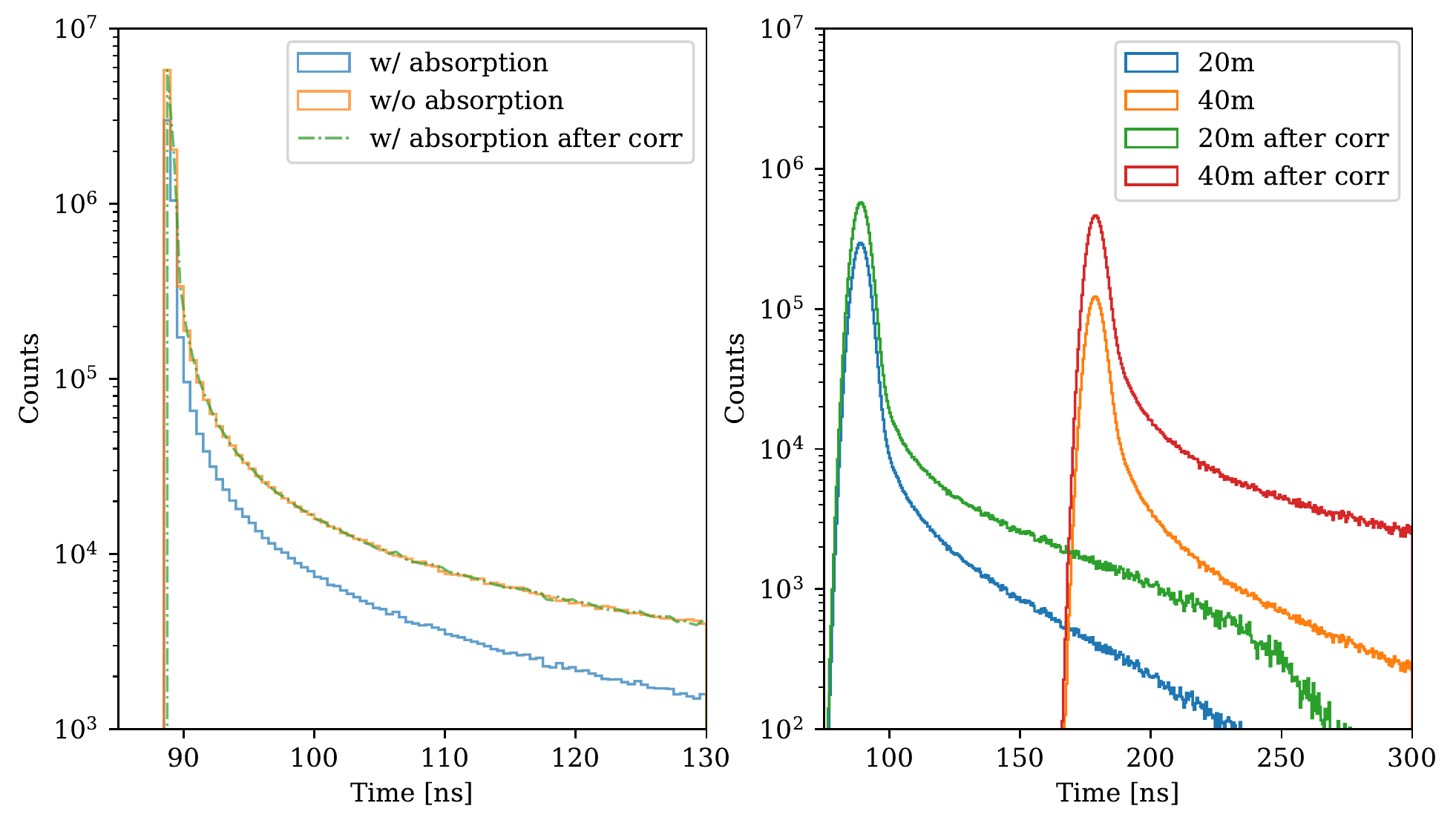}
    \caption{Illustration for measuring absorption length in the PMT system. Left panel: the ATD observed by PMT with and without absorption effect. The PMT is located 20\,m away from the light source, and no time smearing effect is simulated. The distribution with absorption can be perfectly corrected to the distribution without absorption. Right panel: The ATD observed by PMT in 20\,m and 40\,m considering time smearing and absorption effect. The total number of photons after absorption correction (integration of green and red curves) at two distances matches well.}
    \label{fig:pmt_absorption}
\end{figure}

The above method can be verified in the simulation. 
We simulate the ATD with two settings, one with the absorption process and the other without. 
The distribution with absorption can be corrected to the distribution without absorption using the $e^{\frac{c t}{n \lambda_\mathrm{abs}}}$ weighting term, as shown in the left panel of Fig. \ref{fig:pmt_absorption}. 
The comparison between the left and right side of Equation \ref{eq:abs_bin} is shown in the right panel of Fig. \ref{fig:pmt_absorption}, where we have shown the measured ATD at a distance of 20\,m and 40\,m and their absorption corrected distribution. 
The number of photon hits obtained by integrating the ATD at 40\,m and 20\,m distances matches well after the absorption correction, with the former only $\sim 3\%$ smaller than the latter. 
Because we have ignored some late-arrived photons in the analysis since the background noise might cover those photons in the actual experiment.

\section{Discussion}
\label{sec:discussion}

\subsection{Effects of geometry simplification in simulation}

The geometry of the simulation program developed in this work assumes an isotropic symmetry of the light source and extends the receiver shell to increase the simulation efficiency, as shown in Fig. \ref{fig:geometry}. 
While the actual light source is not isotropic due to the penetrator, vacuum port, and other mechanical structures, the data analysis performed always integrates over $\phi$ direction, and the anisotropy in $\theta$ direction can be calibrated in the laboratory and implemented on simulation data. 
The occlusions caused by apparatus such as submarine cable have effects of less than 0.1\% for T-REX and are neglected in the simulation.

The boundary process of photons when going through air-glass and glass-water interfaces are not simulated either. 
While the reflection on the interfaces can reduce the number of photons detector in the photon sensors, our relative measurement strategy can eliminate this effect. 
The effect caused refraction of photons is more complicated than the reflection. 
Since the thickness of the glass shell is uniform, the air-glass and glass-water interfaces can be merged into a single air-water interface when studying the refraction effect. 
Detailed optical analysis shows that while the interface between the emitter and water can shrink the image size of the emitter for an observer in water, the interface between the receiver and water can expand the image size observed by the camera \cite{TianWei:2022}.
So those interfaces do not influence the final image size observed by the camera.

\subsection{Difference between the two systems}

T-REX experiment utilized both the camera system and the PMT system to measure the optical properties of deep seawater.
In our numerical simulation, we constructed various physical quantities from the observables measured by the two systems, as presented in Sec. \ref{sec:validation_of_analysis_methods}. 

The two photon sensors observe the light field at two different distances. 
Their instrumental properties cause the difference in the optical parameters to which they are sensitive. 
The camera system features high angular resolution to the arrived photons. 
The gray values in the image taken by the camera corresponds to the radiance of the steady light field:
\begin{linenomath*}
\begin{equation}
    \mathrm{G}(x, y) \propto I(\theta_\mathrm{in}) ,
\end{equation}
\end{linenomath*}
where $\mathrm{G}$ is the gray value of the image, $(x, y)$ is the coordinate at the image and is related to the incident angle of photons $\theta_\mathrm{in}$ using the geometric relation shown in the lower panel of Fig. \ref{fig:camera_model}. 
The camera system can effectively select directly arriving photons using direction information, i.e., using only the gray values at the center of the image. 
Thus, the camera system is capable of measuring the attenuation length. 

The PMTs can not distinguish photons from different directions and thus can only measure the \textit{irradiance}, which is the integral of radiance over all directions. 
On the other hand, the timing ability enables PMTs to measure the arrival time of the photons from a pulsing light source. 
The PMT system can approximately measure the scattering angle averaged attenuation length by comparing the number of photons received in a time window of $3 \times \mathrm{TTS}$ around the peak at two distances, since the arrival time of those photons indicates that they can not by strongly scattered.
The extra trajectory length induced by the scattering effect can be well calibrated using the arrival time measured by the PMTs. 
Those properties make the PMT system suitable for measuring the absorption length of the deep seawater.

\section{Summary}
\label{sec:summary}

In this paper, we introduce a program based on \textsf{Geant4} toolkit that simulates photon propagation in the deep seawater.
The simulation results are compared to the experimental data taken during the T-REX expedition in South China Sea and show good agreement for both the PMTs and the cameras. 
The program is used to extract the key optical parameters at the selected site, which is crucial for obtaining accurate detector simulation and optimizing detector design of the TRIDENT telescope.

Our simulation provides a thorough examination of different optical parameters. We evaluate the analysis methods using different observables from the PMTs and the cameras. 
The differences between the effective attenuation length - measured by the PMT using photon counts, and the canonical attenuation length - measured by the camera using central gray value of the photographs, were also resolved.
It is shown that the effective attenuation length can get up to $\sim 40\%$ larger than the canonical attenuation length if all photons observed by a PMT are counted.

The simulation program developed can be employed by other experiments with PMTs or cameras to measure optical properties for neutrino telescopes in deep lake water or glacial ice. 
It can also be easily adapted for application in the optical calibration system for the next-generation neutrino telescopes.

\section*{Acknowledgements}

We are grateful to Jun Guo and Iwan Morton-Blake for their help to improve this paper.

This work is supported by the Ministry of Science and Technology of China [No. 2022YFA1605500]; Office of Science and Technology, Shanghai Municipal Government [No. 22JC1410100]; % Xu, Donglian
the Natural Science Foundation of China [No. 11773003, U1931201]; and the China Manned Space Project [No. CMS-CSST-2021-B11]. % Li, Zhuo

\bibliography{bibfile}

\begin{thebibliography}{10}
\expandafter\ifx\csname url\endcsname\relax
  \def\url#1{\texttt{#1}}\fi
\expandafter\ifx\csname urlprefix\endcsname\relax\def\urlprefix{URL }\fi
\expandafter\ifx\csname href\endcsname\relax
  \def\href#1#2{#2} \def\path#1{#1}\fi

\bibitem{AMANDA_overview:2000}
E.~Andres, et~al., {The AMANDA neutrino telescope: Principle of operation and
  first results}, Astropart. Phys. 13 (2000) 1--20.
\newblock \href {http://arxiv.org/abs/astro-ph/9906203}
  {\path{arXiv:astro-ph/9906203}}, \href
  {http://dx.doi.org/10.1016/S0927-6505(99)00092-4}
  {\path{doi:10.1016/S0927-6505(99)00092-4}}.

\bibitem{ANTARES_overview:2011}
M.~Ageron, et~al., {ANTARES: the first undersea neutrino telescope}, Nucl.
  Instrum. Meth. A 656 (2011) 11--38.
\newblock \href {http://arxiv.org/abs/1104.1607} {\path{arXiv:1104.1607}},
  \href {http://dx.doi.org/10.1016/j.nima.2011.06.103}
  {\path{doi:10.1016/j.nima.2011.06.103}}.

\bibitem{icecube_instrument_review:2010}
F.~Halzen, S.~R. Klein, {IceCube: An Instrument for Neutrino Astronomy}, Rev.
  Sci. Instrum. 81 (2010) 081101.
\newblock \href {http://arxiv.org/abs/1007.1247} {\path{arXiv:1007.1247}},
  \href {http://dx.doi.org/10.1063/1.3480478} {\path{doi:10.1063/1.3480478}}.

\bibitem{KM3Net_letter_of_intent:2016}
S.~Adrian-Martinez, et~al., {Letter of intent for KM3NeT 2.0}, J. Phys. G
  43~(8) (2016) 084001.
\newblock \href {http://arxiv.org/abs/1601.07459} {\path{arXiv:1601.07459}},
  \href {http://dx.doi.org/10.1088/0954-3899/43/8/084001}
  {\path{doi:10.1088/0954-3899/43/8/084001}}.

\bibitem{Baikal_status:2011}
A.~Avrorin, et~al., {The gigaton volume detector in Lake Baikal}, Nucl.
  Instrum. Meth. A 639 (2011) 30--32.
\newblock \href {http://dx.doi.org/10.1016/j.nima.2010.09.137}
  {\path{doi:10.1016/j.nima.2010.09.137}}.

\bibitem{Wiebusch:2003}
J.~Ahrens, et~al., {Muon track reconstruction and data selection techniques in
  AMANDA}, Nucl. Instrum. Meth. A 524 (2004) 169--194.
\newblock \href {http://arxiv.org/abs/astro-ph/0407044}
  {\path{arXiv:astro-ph/0407044}}, \href
  {http://dx.doi.org/10.1016/j.nima.2004.01.065}
  {\path{doi:10.1016/j.nima.2004.01.065}}.

\bibitem{IceCube_source_sensitivity:2003}
J.~Ahrens, et~al., {Sensitivity of the IceCube detector to astrophysical
  sources of high energy muon neutrinos}, Astropart. Phys. 20 (2004) 507--532.
\newblock \href {http://arxiv.org/abs/astro-ph/0305196}
  {\path{arXiv:astro-ph/0305196}}, \href
  {http://dx.doi.org/10.1016/j.astropartphys.2003.09.003}
  {\path{doi:10.1016/j.astropartphys.2003.09.003}}.

\bibitem{IceCube-Gen2_layout:2021}
A.~Omeliukh, et~al., {Optimization of the optical array geometry for
  IceCube-Gen2}, PoS ICRC2021 (2021) 1184.
\newblock \href {http://arxiv.org/abs/2107.08527} {\path{arXiv:2107.08527}},
  \href {http://dx.doi.org/10.22323/1.395.1184}
  {\path{doi:10.22323/1.395.1184}}.

\bibitem{ANTARES_measurement:2004}
J.~A. Aguilar, et~al., {Transmission of light in deep sea water at the site of
  the ANTARES Neutrino Telescope}, Astropart. Phys. 23 (2005) 131--155.
\newblock \href {http://arxiv.org/abs/astro-ph/0412126}
  {\path{arXiv:astro-ph/0412126}}, \href
  {http://dx.doi.org/10.1016/j.astropartphys.2004.11.006}
  {\path{doi:10.1016/j.astropartphys.2004.11.006}}.

\bibitem{IceCube_measurement:2006}
M.~Ackermann, et~al., {Optical properties of deep glacial ice at the South
  Pole}, J. Geophys. Res. 111~(D13) (2006) D13203.
\newblock \href {http://dx.doi.org/10.1029/2005JD006687}
  {\path{doi:10.1029/2005JD006687}}.

\bibitem{NEMO_measurement:2006}
G.~Riccobene, A.~Capone, {Deep seawater inherent optical properties in the
  Southern Ionian Sea}, Astropart. Phys. 27 (2007) 1--9.
\newblock \href {http://arxiv.org/abs/astro-ph/0603701}
  {\path{arXiv:astro-ph/0603701}}, \href
  {http://dx.doi.org/10.1016/j.astropartphys.2006.08.006}
  {\path{doi:10.1016/j.astropartphys.2006.08.006}}.

\bibitem{Grace_measurement:2011}
E.~G. Anassontzis, et~al., {Light transmission measurements with LAMS in the
  Mediterranean Sea}, Nucl. Instrum. Meth. A 626-627 (2011) S120--S123.
\newblock \href {http://dx.doi.org/10.1016/j.nima.2010.06.353}
  {\path{doi:10.1016/j.nima.2010.06.353}}.

\bibitem{Baikal_measurement:2012}
A.~Avrorin, et~al., {Asp-15: A stationary device for the measurement of the
  optical water properties at the NT200 neutrino telescope site}, Nucl.
  Instrum. Meth. A 693 (2012) 186--194.
\newblock \href {http://dx.doi.org/10.1016/j.nima.2012.06.035}
  {\path{doi:10.1016/j.nima.2012.06.035}}.

\bibitem{Grace_measurement:2018}
K.~G. {Balasi}, D.~{Lenis}, M.~{Maniatis}, N.~{Maragos}, G.~{Stavropoulos}, {A
  method for measuring the optical parameters of deep-sea water}, Frontiers in
  Physics 6 (2018) 132.
\newblock \href {http://dx.doi.org/10.3389/fphy.2018.00132}
  {\path{doi:10.3389/fphy.2018.00132}}.

\bibitem{P-One_measurement:2021}
N.~Bailly, et~al., {Two-year optical site characterization for the Pacific
  Ocean Neutrino Experiment (P-ONE) in the Cascadia Basin}, Eur. Phys. J. C
  81~(12) (2021) 1071.
\newblock \href {http://arxiv.org/abs/2108.04961} {\path{arXiv:2108.04961}},
  \href {http://dx.doi.org/10.1140/epjc/s10052-021-09872-5}
  {\path{doi:10.1140/epjc/s10052-021-09872-5}}.

\bibitem{TRIDENT_arxiv:2022}
Z.~P. Ye, et~al., {Proposal for a neutrino telescope in South China Sea}\href
  {http://arxiv.org/abs/2207.04519} {\path{arXiv:2207.04519}}.

\bibitem{Wenlian:2022}
W.~Li, et~al., The Light Source of the TRIDENT Pathfinder Experiment, in
  preparation.

\bibitem{GEANT4:2003}
S.~Agostinelli, et~al., {GEANT4--a simulation toolkit}, Nucl. Instrum. Meth. A
  506 (2003) 250--303.
\newblock \href {http://dx.doi.org/10.1016/S0168-9002(03)01368-8}
  {\path{doi:10.1016/S0168-9002(03)01368-8}}.

\bibitem{GEANT4:2006}
J.~Allison, et~al., {Geant4 developments and applications}, IEEE Trans. Nucl.
  Sci. 53 (2006) 270.
\newblock \href {http://dx.doi.org/10.1109/TNS.2006.869826}
  {\path{doi:10.1109/TNS.2006.869826}}.

\bibitem{book_for_optical:1998}
C.~F. Bohren, D.~R. Huffman, Absorption and Scattering of Light by Small
  Particles, John Wiley \& Sons, Ltd, 1998, Ch.~1, pp. 1--11.
\newblock \href {http://dx.doi.org/https://doi.org/10.1002/9783527618156.ch1}
  {\path{doi:https://doi.org/10.1002/9783527618156.ch1}}.

\bibitem{Henyey_Greenstein:1941}
L.~G. {Henyey}, J.~L. {Greenstein}, {Diffuse radiation in the Galaxy.},
  Astrophysical Journal 93 (1941) 70--83.
\newblock \href {http://dx.doi.org/10.1086/144246} {\path{doi:10.1086/144246}}.

\bibitem{Mayerhofer:2020}
T.~G. Mayerh{\"o}fer, S.~Pahlow, J.~Popp, The bouguer-beer-lambert law: Shining
  light on the obscure, ChemPhysChem 21~(18) (2020) 2029--2046.
\newblock \href {http://dx.doi.org/10.1002/cphc.202000464}
  {\path{doi:10.1002/cphc.202000464}}.

\bibitem{Vountas:2003}
M.~Vountas, A.~Richter, F.~Wittrock, J.~Burrows, Inelastic scattering in ocean
  water and its impact on trace gas retrievals from satellite data, Atmospheric
  Chemistry and Physics 3~(5) (2003) 1365--1375.

\bibitem{KM3Net_mDOM_simulation:2016}
C.~M.~F. Hugon, {GEANT4 simulation of optical modules in neutrino telescopes},
  PoS ICRC2015 (2016) 1106.
\newblock \href {http://dx.doi.org/10.22323/1.236.1106}
  {\path{doi:10.22323/1.236.1106}}.

\bibitem{IceCube-Gen2_sensor:2021}
N.~Shimizu, et~al., {Performance studies for a next-generation optical sensor
  for IceCube-Gen2}, PoS ICRC2021 (2021) 1041.
\newblock \href {http://arxiv.org/abs/2108.05548} {\path{arXiv:2108.05548}},
  \href {http://dx.doi.org/10.22323/1.395.1041}
  {\path{doi:10.22323/1.395.1041}}.

\bibitem{JUNO_simulation:2018}
T.~Lin, J.~Zou, W.~Li, Z.~Deng, G.~Cao, X.~Huang, Z.~You, {Parallelized JUNO
  simulation software based on SNiPER}, J. Phys. Conf. Ser. 1085~(3) (2018)
  032048.
\newblock \href {http://arxiv.org/abs/1710.07150} {\path{arXiv:1710.07150}},
  \href {http://dx.doi.org/10.1088/1742-6596/1085/3/032048}
  {\path{doi:10.1088/1742-6596/1085/3/032048}}.

\bibitem{ROOT:2011}
I.~Antcheva, et~al., {ROOT: A C++ framework for petabyte data storage,
  statistical analysis and visualization}, Comput. Phys. Commun. 182 (2011)
  1384--1385.
\newblock \href {http://dx.doi.org/10.1016/j.cpc.2011.02.008}
  {\path{doi:10.1016/j.cpc.2011.02.008}}.

\bibitem{TianWei:2022}
W.~Tian, et~al., An optical property calibration system based on CMOS camera
  for deep sea neutrino telescopes, in preparation.

\bibitem{Yudi:2022}
F.~Zhang, et~al., Characterize the PMT System in the TRIDENT Pathfinder
  Experiment, in preparation.

\end{thebibliography}

\end{document}